# Designing L5:
# A Permacomputing Approach to Creative Coding

Lee Tusman
lee.tusman@purchase.edu
New Media/Math & Computer Science
State University of New York at Purchase
Purchase, NY USA

Kit Kuksenok
kit@processingfoundation.org
Processing Foundation
p5.js Project Lead
Berlin, DE

## ABSTRACT

Creative coding libraries provide high-level tools that make computational and algorithmic art accessible to artists and learners. Processing/p5 is one such family of libraries, known for its beginner-friendly approach and wide reach across artistic and technical communities. L5 is a new member of this family, implemented in Lua using the LÖVE framework. It applies permacomputing principles, a movement addressing sustainability in computing inspired by permaculture, bringing these values to a community of practice not historically centered on them. This paper explores L5's design decisions and tensions between sustainability and usability through five case studies: 1. balancing perceived simplicity versus exposing the seams, 2. designing for lower resource consumption, 3. ensuring long-term stability, 4. constraining functionality, and 5. designing documentation for resource-constrained access. Rather than optimizing for a single metric, sustainable creative tools require navigating competing values transparently.

## 1 INTRODUCTION

Creative coding is both an artistic practice and a site for critical engagement with computers. We look at how this critical lens could, through the tools used, be turned toward sustainability and resource awareness. Creative coding communities have historically largely not centered permacomputing or sustainability values. Our goal is not efficiency in the managerial sense but a more situated resource awareness, attentive to machines at hand and the context of computational art-making [13].

Processing/p5 is widely known and offers significant potential for bringing permacomputing principles and practices to creative coding communities. Personal computing is now broadly recognized as extractive and unsustainable, yet computers remain an essential and pervasive part of daily life. The question L5 asks is what tools can empower people within the frame of artistic expression and experimentation, while centering sustainability, reduced resource consumption, and longevity.

### 1.1 Creative Coding Context

Processing describes itself as a "flexible software sketchbook and language for learning how to code [19]." Created by Casey Reas and Ben Fry at MIT's Aesthetics + Computation Group, Processing consists of a language library implemented in Java and an Integrated Development Environment (IDE). Since its release in 2001, Processing has grown into a family of implementations across multiple programming languages, stewarded by the Processing Foundation and sustained by a global community of artists, educators, and students.

### 1.2 Processing as a Family of Languages

Processing/p5 is a set of shared computational and community practices implemented across programming languages, with millions of users over the past two decades [12]. We use the term Processing/p5 to name these implementations without specifying the underlying language.

The original Processing Java implementation allowed exporting applets for the web until Java was phased out from browsers. Processing.js (2008) enabled Processing code to run in browsers by converting Java syntax to JavaScript. In 2013, Lauren McCarthy initiated p5.js, a fundamental reimplementation of Processing's API in native JavaScript designed for web-based creative coding. Beyond Java and JavaScript, Processing/p5 libraries have emerged in other languages: Quil in Clojure/ClojureScript, JRubyArt in Ruby, and py5 in Python, among others. Each implementation adapts Processing/p5 design philosophy to its host language while maintaining the core principles. L5 (Fig. 1) extends this family by bringing Processing/p5 to Lua [25], a lightweight language particularly suited to embedded systems and resource-constrained environments [8]. L5 does this by building on LÖVE, a free and open source game framework for Lua that handles graphics, audio, and input. The L5.lua single-file library implements Processing/p5 conventions and drawing paradigm on top of LÖVE, and can be used in a text editor of the user's choosing.

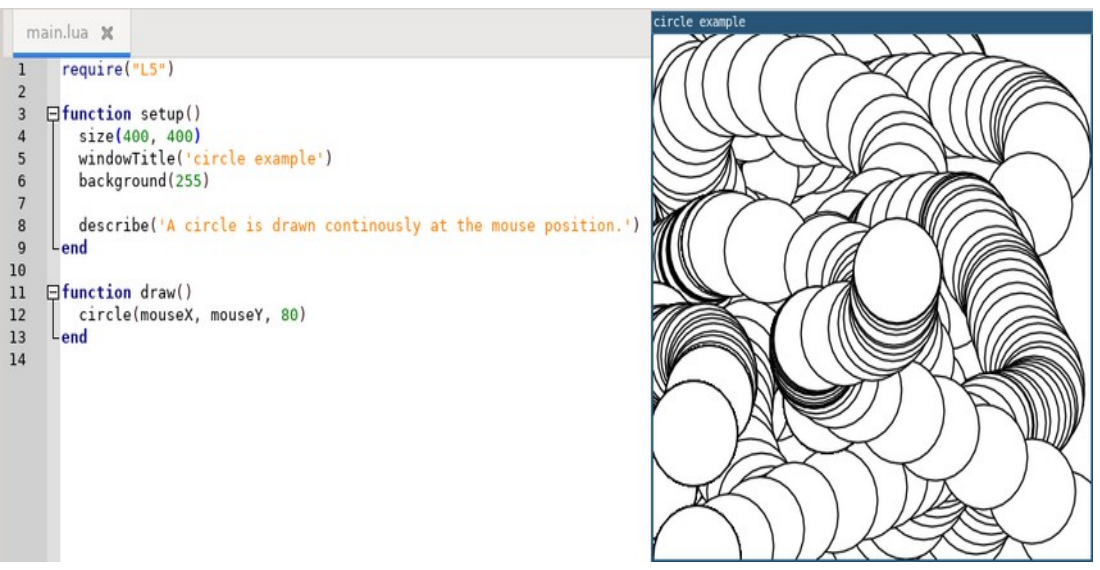

Figure 1: **L5 code sketch written with L5 library in Geany editor, with running program output after mouse interaction. A circle is drawn continuously with mouse interaction.**

### 1.3 Design Philosophy: From Access to Sustainability

Just as p5.js centers accessibility through its Access Statement [17], L5 extends this practically by centering sustainability through permacomputing principles [23]. The p5.js Access Statement makes a commitment to center the needs of people who are marginalized or excluded by dominant computing culture by only considering those features that increase access and inclusion. Older hardware is included as one dimension of access [17] among several others. L5 asks, specifically: Does this feature work on older hardware? Does it minimize resource consumption? Will it remain functional long-term? Does it reduce barriers to participation for users with constrained or aging systems? L5 builds on p5.js's values, treating the continued viability of older computers as creative computing platforms as an extension of the accessibility commitments p5.js already articulates.

### 1.4 Permacomputing Principles

Permacomputing is a concept and growing movement examining sustainability in computing, evolving from related fields including collapse computing, sustainability, right to repair, FLOSS, low-tech and others [4]. It has grown into a community of practice organized through meetups in Europe, the United States, as well as through online forums including an email list and collaboratively-maintained wiki. This community encompasses multiple approaches, from considering the lifecycle of computing machinery to intersectional and anti-capitalist critique of technology, inviting computer users "to collectively and radically rethink computational culture [26]."

L5 is one response to this invitation, growing out of the goal of one artist and educator to "dig where we stand" [6] with Processing/p5 while rethinking its goals and implementation with permacomputing principles as a guide. The pedagogical intent of L5 making these principles explicit is to offer artists, educators and students a practical entry point into thinking about sustainability and resource limits in computing. Introducing limits-aware values into technical communities is a recognized challenge [2,18] and L5 approaches this through educational tool design and learner-oriented documentation.

The permacomputing wiki suggests starting points such as helping a school work with recycled computers; researching how to provide less resource-intensive tools and systems for one's lab or workspace; or helping artists to engage with ecological topics using tools and media in line with this intention. L5 addresses these areas, and intentionally incorporates permacomputing into the context of creative coding.

This paper presents a reflexive account of the design decisions made in developing L5, identifying five tensions that emerged from applying permacomputing principles to a creative coding library. Each case study that follows presents a tension point, the specific design decision taken with the relevant permacomputing principle, and a discussion of its implementation in context. Rather than evaluating L5 against a predetermined framework, we focus on tensions emerging through the design and implementation process itself, and reflect on moments when we found permacomputing values pulling against usability and familiarity.

The authors are artists and educators. One author organizes permacomputing community meetups, leads L5's development, and makes artwork with it; the other is the current p5.js project lead. L5 is a collaborative open source project shaped by over a dozen code and documentation contributors and a broader community drawn from both creative coding and permacomputing, through GitHub issues, usability testing, and meetups. This paper is written from the perspective of L5: examining design decisions reflexively rather than from an external standpoint, and presenting both current capacities of L5 as a software, and our aspirations for it as a community of practice.

# 2 CASE STUDY 1: BALANCING PERCEIVED SIMPLICITY AND EXPOSING THE SEAMS

## 2.1 Tension Point

Creative coding libraries can appear "simple" in different ways, with different hidden costs. The Processing Java library comes bundled with a beginner-friendly IDE that abstracts away Java's complexity. The p5.js Web Editor offers immediate, zero-install coding but depends on cloud infrastructure (such as Amazon Web Services), browser compatibility, and continuous deployment that supports robust maintenance by contributors, including newcomers to open source and coding. This storage footprint and computational cost is invisible to users. Each simplification for users, including seamless editors, automatic updates, and managed dependency toolchains, carries hidden computational costs.

## 2.2 Decision

Following the permacomputing principle to Expose the Seams, L5 generally chooses transparency over seamlessness, making its dependencies and toolchain explicit. Its toolchain must be assembled by the programmer to set up their coding environment. Cloud infrastructure reliance is generally minimized. The L5 library is currently a single file of under 5000 lines of code, kept deliberately small so that its implementation remains legible to users. The seam remains, but is intentionally kept manageable.

## 2.3 Context

There are different relevant, sometimes competing, definitions of simplicity with respect to developer experience. Many languages could be considered simple in particular ways, such as implementations of Forth, which attempts to eliminate the “vast hierarchy of languages” into a single layer bridging low-level and high-level programming contexts [20]. L5 is a high-level general-purpose creative library attempting to minimize computational impact while remaining accessible and usable by beginners, artists, designers, and students.

The most computationally minimal choice of software stacks is rarely the most beginner-friendly, and features that support beginners (helpful error messages, extensive tooling, larger API) consume resources. L5's development navigates these trade-offs on a case by case basis.

When evaluating creative coding environments, we distinguish between perceived simplicity—what appears easy to users, and computational simplicity, the actual computational and infrastructural costs. As the Permacomputing principles state, "what is perceived as simple can be energy inefficient and arcane [27].” This distinction becomes clear when examining the complete lifecycle of creative coding work: setup, development, sharing, archiving, and maintenance.

Maintenance of creative coding libraries reflects accumulated complexity hidden by perceived simplicity. As Michael Murtaugh critiques in "Torn at the seams," even "seamless" environments like Processing involve hidden abstractions that shape creative practice in particular ways [16]. L5 makes different design decisions based on this choice to expose more of the seams hidden by perceived simplicity. For example, the p5.js Web Editor offers immediate gratification: you write code in the browser, see results instantly, and store to the cloud. This seamless experience depends on continuous server availability, internet connectivity, account management systems, and client-server synchronization. Code changes transmit data and project saves are written to cloud storage. Computational cost is externalized and invisible.

The p5.js Web Editor requires no local maintenance by users but depends on external infrastructure. Breaking changes as a result of updates to the 2.0 version of the p5.js library itself require code updates, which the Web Editor now makes visible. Users explicitly select which version of p5.js their project runs on. Processing and L5 operate locally, with different trade-offs. Processing's PDE (Processing Development Environment) provides an integrated experience similar to the p5.js Web Editor but runs entirely on a user's machine. L5 requires users to manage files manually and run programs through LÖVE, making the toolchain visible but taking care of dependencies.

In addition to Processing/p5 languages, we consider game engines such as Unity and Unreal as additional related toolkits for engaging in programming and artistic creation. These systems normalize use of large hard drives, cloud storage and version control through their ecosystems. Unity and Unreal Engine require continuous maintenance to keep up with engine updates, platform changes, and graphics API evolution. These workflows require persistent internet connectivity, recent computers and graphics cards, cloud service reliance, and significant bandwidth for file syncing.

Processing requires periodic updates to address operating system and Java changes, but projects can be maintained locally without updating. L5 maintenance is explicit and minimal: Lua stability and LÖVE's conservative development philosophy mean projects require infrequent updates. When updates are necessary (e.g., LÖVE major version changes), these are explicit events that happen every few years, announced well in advance. The trade-off is that users manage their own installations, though they can remain on older versions indefinitely without breakage, and that the scale and type of projects that can be developed with a tool like L5 or Processing is different from what can be developed with Unity or Unreal Engine.

Lifecycle differences have pedagogical implications. What do different environments teach beginners about computation, infrastructure, and sustainability? p5.js is a particularly beginner-friendly library. It can be launched ready to go from the p5.js website directly to the web editor without needing to install anything, to log in, or to understand the interplay of web technologies for basic functionality. The p5.js Web Editor's seamless experience allows a student to get on with their work without necessarily considering their use of the cloud, how their work is shared through URLs or the maintenance work behind the scenes.

Processing and L5 both make a user’s own system more visible to them: including directory and file structure, application launching, and library dependencies. This visibility has a steeper learning curve for beginners, with the benefit of a more accurate mental model of how software works.

From a permacomputing perspective, L5 teaches sustainability alongside creative coding by making dependencies and system requirements explicit, keeping storage requirements minimal, and operating offline-first. The "friction" of assembling a toolchain becomes an opportunity to understand what software requires.

The absence of cloud infrastructure is not necessarily a missing feature but a factor of its lifecycle: creative coding does not inherently require internet connectivity, server infrastructure, or automatic updates.

This approach trades perceived simplicity (no seamless web editor, manual toolchain assembly, visible file management) for computational simplicity (minimal storage, no cloud dependencies, long-term stability, explicit costs). Whether this trade-off serves creative coders depends on context: for workshops in environments with limited internet connectivity, L5's offline-first design becomes enabling rather than limiting. For casual experimentation, p5.js's zero-friction entry point may be more appropriate. L5 positions itself working offline, less reliant on robust cloud infrastructure, and focusing on projects with potential long-term viability.

The sharing lifecycle reveals more tensions between perceived and computational simplicity. p5.js offers one-click sharing: projects “saved” in the p5.js Web Editor are hosted and immediately accessible via URL. This convenience depends on the Processing Foundation's continued maintenance of hosting infrastructure and assumes recipients have modern browsers and internet access.

Processing applications can be exported as standalone executables for different platforms, but may require users to understand platform-specific packaging and increase file sizes significantly (each exported application bundles the Java runtime). Unity and Unreal have similar export complexity with even larger file sizes.

L5 projects can be distributed in two ways. For users with LÖVE installed, projects are shared as folders containing Lua files and the L5.lua library, forming a small program portable via USB drive, local network, or any file transfer method without requiring internet infrastructure. Alternatively, L5 projects can be packaged as standalone executables using LÖVE's built-in distribution tools, bundling the LÖVE runtime with the project code. This creates self-contained applications similar to Processing exports but with significantly smaller file sizes due to LÖVE's minimal footprint compared to bundling the entire Java runtime. Both distribution methods work entirely offline.

## 3 CASE STUDY 2: DESIGNING FOR LOWER RESOURCE CONSUMPTION

### 3.1 Tension Point

A tool with a larger footprint in storage, computation, and bandwidth can expand what artists are able to make and smooth the user experience. In Case Study 1 we considered how a higher footprint can serve perceived simplicity; here we consider the resource impact of the features a tool makes available. Processing's Java foundation requires a substantial memory baseline even for simple sketches. p5.js depends on browser rendering engines that grow more resource-intensive with each generation. Game engines like Unity and Unreal Engine, sometimes adapted for creative coding, escalate these demands further, requiring dedicated graphics hardware, gigabytes of storage, and frequent updates that make older installations obsolete. The promise of these more powerful environments is richer real-time graphics, more complex simulations, and a broader feature set. Lighter tools preserve access and longevity, but they also impose constraints that can themselves drive creative innovation.

### 3.2 Decision

Following the permacomputing principles of Care for All Hardware and Not Doing, the hardware demands, energy use, and e-waste associated with feature-rich tools are not unfortunate side effects but the direct result of design decisions that reproduce their own implicit feature and capability expectations with no room for critical consideration. L5's design attempts to remain functional on hardware that other creative coding environments have already left behind. It becomes a design value that expands who can participate in computational art-making across economic and access contexts.

### 3.3 Context

Permacomputing-oriented software projects can take many forms: reducing energy consumption, extending hardware lifespans, minimizing dependencies, or resisting planned obsolescence. Within permacomputing discourse, particular concern centers on software bloat (unnecessary feature accumulation that increases resource requirements), software rot (degradation of functionality over time due to dependency changes), and the escalating hardware requirements of contemporary software that render older machines obsolete. L5 addresses these concerns through specific design choices aimed at reducing bloat, addressing longevity and maintaining broad compatibility [14].

L5 prioritizes working within constraints. The Permacomputing wiki describes software bloat as "long, slow or otherwise wasteful of resources," referencing Wirth's law: "software is getting slower more rapidly than hardware is getting faster [24].” The entire L5 library is under 200 KB, along with the LÖVE framework dependency it is 5 MB on Linux, and up to 25MB on macOS, including the Lua language. L5 depends only on Lua and LÖVE, both stable projects with long maintenance histories, avoiding the complex dependencies common in modern software ecosystems. Programs run efficiently on hardware tested from the past 10 - 15 years, and no internet connectivity is required for development or deployment.

An initial barrier to participating in creative coding varies dramatically across environments. Unity requires 1-3GB per installed version, and Unreal Engine requires dozens of gigabytes or up to 100GB with all modules. While not designed specifically for creative coding, their utility for creating 2D and 3D environments makes them common choices not only for developing games but also for building simulations, interactive visualizations and immersive environments [15,21]. These environments demand recent, powerful computers with significant graphics processing capability and hard drive space.

Processing appears more accessible: approximately 468MB (Processing 4.3.2 for Linux, retrieved 2025-02-02 from processing.org) including its Java runtime. Other Processing/p5 implementations vary: py5 requires Python3 (25-100MB) plus Java (50-300MB); Quil requires the Java Virtual Machine as well, plus the Leiningen build tool (approximately 10MB).

p5.js presents the most accessible entry point: the library itself ranges from under 1MB minified to several megabytes with additional libraries. However, this apparent simplicity obscures other system requirements: users require a current web browser (250-450MB for popular commercial browsers), and the p5.js Web Editor, in addition to the computational and maintenance costs of server infrastructure, continuous deployment systems, and browser requirements.

L5's total storage footprint (LÖVE + L5.lua + minimal editor) ranges from approximately 5MB to 45MB depending on platform and editor choice. This is orders of magnitude smaller than Unity/Unreal, substantially smaller than Processing, and similar to or much smaller than p5.js when including browser requirements.

LÖVE uses LuaJIT by default, a widely-used just-in-time compiler based on Lua 5.1, rather than the standard PUC Lua interpreter. This matters for resource consumption: as the authors of The Green Side of Lua explain, "programming languages and execution models strongly influence software energy consumption." In this preprint, the authors find that LuaJIT consumes roughly seven times less energy while running seven times faster than the best standard Lua interpreter, approaching C's efficiency [3]. LÖVE's use of JIT compilation combines the speed benefits of compiled languages with the flexibility and immediate feedback of an interpreted, dynamically-typed language. For L5, this means a high-level, beginner-friendly syntax does not come at the cost of performance. L5 can produce fast, efficient code even on resource-constrained hardware.

Lua’s design constraints and those of LÖVE were selected because they support a focus on lightweight computing. Lua was designed to support running on a variety of hardware including embedded systems. It has a small memory footprint and uses a minimal syntax and minimum of data structures (the table, for example) to implement features. Compared to a language like Python, it has a much smaller functionality built-in, but this tradeoff was selected in order to support its other features [7].

# 4 CASE STUDY 3: DESIGN FOR LONGEVITY

## 4.1 Tension Point

Designing a creative coding library for longevity means ensuring both that the tool itself remains viable over time and that programs made with it can be run years or decades after they were written. Long-standing open-source projects are also committed to longevity, though continuous updates and evolving tooling can become a source of instability, introducing breaking changes. The alternative is equally fraught: minimizing dependencies or freezing development risks breakdown as underlying hardware and operating systems shift, or as new security problems emerge. Longevity as a design goal does not resolve this tension so much as reframe it, trading the churn of continuous feature growth for the slower challenge of remaining functional and useful across time.

## 4.2 Decision

Following the permacomputing principle of Build on Solid Ground, L5 treats longevity as a deliberate design commitment. It minimizes dependencies in its build system, and grounds itself in slowly-evolving, long-maintained tooling: Lua, LuaJIT, and LÖVE, each notable for their emphasis on long-term stability. Future hardware cannot be tested, but the past can: L5 is tested on machines as far back as possible, using older hardware as a proxy for long-term viability. This backward compatibility testing is an imperfect but concrete commitment to the longevity goal and a way of operationalizing an aspiration that can never be fully verified in advance.

## 4.3 Context

Permacomputing concerns itself with computing technology's impact on the environment, taking inspiration from permaculture. As planned obsolescence for phones, tablets, and computers is part of the dominant Silicon Valley model, L5 attempts to support as wide a spectrum of devices as possible for as long as possible, supporting creative coding that doesn't require constant hardware upgrades, and expanding access for those with older or more resource-constrained machines.

Lua has been continuously developed by a small team at PUC-Rio in Brazil and is slowly, intentionally updated [10]. It is common in the Lua community to write software that runs on older versions of Lua — in contrast to Python's breaking changes, JavaScript's monthly updates, or the commercially driven evolution of Java. LÖVE, begun as an open source game library in 2008, has been maintained and documented in dialog with its community ever since. LÖVE uses LuaJIT by default, an alternative Lua runtime that is significantly faster than standard PUC Lua [9], further supporting performance on older hardware.

To archive a p5.js project for long-term preservation, one would have to download the p5.js library and sketch file, but the browser itself presents a significant challenge: preserving a browser environment across decades is difficult and imperfect. Continual viability of p5.js code also relies on the canvas element as its underlying framework, and the JavaScript ecosystem is subject to continual change, with updates shipped independently by competing browser manufacturers, creating ongoing compatibility challenges [5]. Processing faces different archival challenges as the Java ecosystem and operating systems evolve, requiring ongoing community efforts to maintain compatibility. As Jang et al. note, hardware and software degradation together jointly limit how long devices remain usable [11].

L5 projects are plain text Lua files referencing a single library file. Lua has changed infrequently over decades, with version 5.0 released in 2003 and 5.5 in 2025 [10]. LÖVE updates are similarly conservative, allowing projects to remain viable across versions. L5 programs have successfully run on hardware as old as an Eee PC 1005HA from 2009, demonstrating backward compatibility across nearly two decades of hardware evolution. While continued LÖVE maintenance is not guaranteed, 18 years of active development and a stable community suggest reasonable prospects for longer-term viability.

## 5 CASE STUDY 4: CONSTRAINING FUNCTIONALITY

### 5.1 Tension Point

As L5 is a Processing/p5 creative coding library, users will naturally expect it to behave like Processing and p5.js. But fully matching their APIs would require extensive native implementations or substantial dependencies, increasing resource consumption and maintenance burden. A smaller, more constrained API keeps the codebase lean, reduces dependencies, and stays viable on older hardware over time. Yet it risks losing the familiarity and expressive range that makes Processing/p5 useful as a creative coding foundation.

### 5.2 Decision

Balancing Processing/p5’s general-purpose graphics and creative tooling with an emphasis on sustainability and longevity means that at times choices must be made. Following the permacomputing principles of Not Doing and Consider Carefully the Interaction Between Simplicity, Complexity and Scale, functionality was only implemented if it could be built through native LÖVE or Lua without external dependencies. For this reason, external libraries beyond LÖVE itself are not leveraged for core L5 library functionality. L5’s API contains over 200 core functions compared to p5.js's 450+ and Processing’s 500+, reducing maintenance burden for the project. These constraints deliberately limit L5's scope: at present no 3D graphics, no web deployment, and a smaller feature set. But they enable functionality to work reliably across diverse contexts: older laptops, resource-constrained environments, installations requiring long-term stability, and situations with limited connectivity.

### 5.3 Context

Programming languages and libraries can be understood as layers of abstractions built on lower-level systems. Processing wraps Java and the Java Virtual Machine, while p5.js is implemented in JavaScript. These are not low-level languages themselves but rather extensions that provide domain-specific functionality—in Processing's case, graphics primitives and creative coding patterns—built on general-purpose language foundations.

L5 follows this pattern: LÖVE handles low-level concerns like graphics rendering, audio, and input through bindings to underlying C libraries. L5 provides the Processing-style API layer, `setup()`, `draw()`, `ellipse()`, `rect()`, that creative coders expect, while Lua and LÖVE handle the computational heavy lifting.

Unlike Forth or C, which demand expert domain knowledge and require programmers to manage low-level graphics libraries directly, Lua offers a relatively simple syntax familiar to users of JavaScript or Python. It is deliberately minimal with straightforward semantics. However, LÖVE core functionality in some cases must be over-ridden to allow a Processing-like approach to stateful programming and the Processing event loop. Whereas LÖVE teaches and restricts the programmer to use `love.load()` to set up sketch defaults, and separates continuous functions to `love.update()` for handling processing and math and physics computation, and `love.draw()` for the rendering phase of a program loop, Processing/p5’s model allows drawing graphics in all of these functions and adds event functions not present in the default LÖVE handling.

Open source software encourages this kind of re-making of software to suit new functionality. The well-documented LÖVE, with the ability to alter its core functionality and event loop through modifying `love.run()`, `love.load()`, and `love.draw()` are key elements that allow for its fluidity to support a different paradigm. Working against some of the speed-ups of Lua/LuaJIT from re-working LÖVE compromises efficiency, but the already-efficient language and runtime helps to offset these changes. In this case, we considered the cost worth the benefit of increased access.

Within the first week of L5’s implementation, dozens of functions were implemented. As Lua, like JavaScript, is a weakly typed scripting language, many of the L5 functions were modeled more closely to p5.js than the Processing-Java equivalent. Whereas p5.js with its emphasis on the browser as platform includes web functionality, no web functionality was implemented in L5 as it targets the native desktop. Additionally, as p5.js has added helper functions and accessibility features, L5 includes initial implementation of `describe()` for alt text to be included in the command line, for use by screen readers. L5 also allows specifying color via HTML color names, like p5.js, implemented through a lookup table.

Processing and p5.js each contain additional bundled renderers for 3d graphics library functionality, including modeling and lighting. A 3d rendering library would increase L5's scope considerably, with raytracing, lighting models or additional modules to be implemented. There are several open source 3d libraries that can be considered but would expand maintenance scope. In L5, these were deferred to a later date to decide, in community, whether and how to implement them, potentially as optional add-on libraries.

Implementing Processing's `filter()` function in L5 required shader code for rendering speed, enabling more complex graphical output but requiring a capable graphics processor. After L5's alpha release, a bug report from a user on a 15-year-old laptop prompted the acquisition of several machines from 2009, 2012, and 2014 for testing. Rather than declaring older hardware unsupported, the shader code was rewritten to work on older GPUs, with fallbacks added so that missing capabilities would notify users with a warning rather than prevent their program from running.

Some API functionality is currently unimplemented due to the requirement for adding in additional dependencies that would increase the size and scope of L5. For example, although JSON is native to JavaScript and widely used in p5.js, supporting it in L5 would require an additional parsing library. However, loading and saving data files as Lua tables in CSV and Lua data was possible to implement using native Lua and LÖVE functionality.

In keeping with its lightweight design, L5 does not currently include its own IDE. To remain accessible to learners, beginners, artists and designers less inclined to use the command line, the `printToScreen()` function was added to display `print()` output natively within programs. This function emerged from L5's first contributor meeting, as part of a broader discussion about easing installation and use for a wider community.

Creators working with imported media such as images and video are also subject to constraints in L5 imposed by LÖVE, which supports JPEG, PNG, BMP, and similar formats but not WebP, and Ogg Theora for video but not common formats like MP4. L5 inherits these limitations and must clearly document them for its users, along with guidance on converting files to supported formats. L5 treats this as an opportunity: learning to convert between file formats and understand their affordances is itself a form of computational literacy, for example through documentation such as the reference and tutorials.

# 6 CASE STUDY 5: WEBSITE INFRASTRUCTURE – THE NO-JAVASCRIPT, LIGHTWEIGHT, OFFLINE-FRIENDLY DOCUMENTATION SITE

## 6.1 Tension Point

Documentation sites are the primary point of introduction, download, reference, and learning for a programming library. Most are built with complex toolchains, heavy JavaScript dependencies, and design systems that prioritize visual polish, assumptions that break down for users on limited bandwidth, older browsers, or unreliable internet connections. Interactive features like in-browser code editors and dynamic search are increasingly standard, but carry costs: larger page weights, JavaScript requirements, and dependence on continued server infrastructure. For L5, whose goals include running on older hardware and remaining viable long-term, documentation that cannot be accessed under the same constraints as the library itself would be a contradiction. The challenge is building something useful, navigable, and attractive without reproducing the resource assumptions L5 is designed to avoid.

## 6.2 Decision

Following the permacomputing principles of Expose the Seams and Care for All Hardware, the website strikes a balance between attractive aesthetics and functionality. L5's documentation site is built with Mkdocs documentation static site generator, chosen so that the underlying content could be migrated to a different generator with little effort. All pages are under 1MB including media, which uses the compressed but widely-available WebP format, and animated GIFs where displaying motion aids understanding. The site is tested on a variety of current, older, low-bandwidth and text-only browsers, served over both HTTP and HTTPS. Functionality prioritizes what static HTML can provide, with optional JavaScript to enable site search. The website is also designed to be downloaded whole, with a text-only and an images-included version available for offline use, re-distributable peer-to-peer or via USB, consistent with how L5 projects themselves can be shared.

## 6.3 Context

A static site generator was selected for its templating, portability, and ability to be hosted anywhere without a database. After evaluating options including those used by Processing and p5.js, MkDocs was chosen as the best fit. It supports viewing without JavaScript, needed only for optional site search. It is beginner-friendly, allowing markdown pages and easy hierarchical organization, and is free and open source. The Download page of the website also includes downloadable copies of the site, available with images or text-only.

This approach has precedent in the LIMITS community, where Abbing demonstrates how static, low-bandwidth design can serve as a principled response to the resource costs of conventional web infrastructure [1]. Following a similar principle, the L5 site does not implement automatic HTTPS redirects, dual-serving HTTP and HTTPS to avoid excluding users on older hardware or software [22]. Reference pages range from 25kB to approximately 250kB depending on the number of animated examples included.

The site was tested on as many browsers and computers as possible, including major browsers and their derivatives, lightweight browsers NetSurf and Dillo, and text-based browsers w3m, links, and offpunk. Images use WebP and animated GIF formats, individually optimized for lowest file size while maintaining adequate fidelity, and are lazy-loaded to reduce resource usage until needed. The Download page includes copies of the site available with images or text-only for offline use.

There is a tradeoff between small image size and fidelity. For WebP, we attempted the smallest possible size from compressing screenshots of L5 sketches running. For gif, video recordings of running sketches were captured as mp4, then converted to gif and compressed at various sizes through imagemagick and ezgif, then individually selected for best possible fidelity at lowest file size. Images are lazy-loaded, reducing resource usage until ready to be viewed by a site visitor.

Building a documentation site that reflects permacomputing principles differs significantly from typical design paradigms. The site treats text and information as first-class, minimizes JavaScript, and avoids extraneous graphics and fonts beyond a small, widely compatible font with native browser fallbacks. Testing on older machines, including a 2009 Windows computer, revealed both successes and areas for improvement. User feedback including appreciative comments, issues, and pull requests confirmed that these design decisions resonated with the community.

## 7 IMPLICATIONS AND FUTURE WORK

L5 is less than a year old and currently in alpha. The design decisions documented here are early ones, and some may prove misguided or require revision as the project matures and finds users. The tensions identified in these case studies point toward open questions rather than settled answers, even while users and contributors have begun documenting their projects with L5 online. Backward compatibility testing on older hardware is ongoing; expanding the range of tested machines and operating systems and documenting this would strengthen L5's longevity claims. The documentation site's offline distribution model could be developed further, including more formal support for peer-to-peer sharing workflows. API coverage relative to Processing and p5.js remains incomplete, and future work includes evaluating which missing features can be added within L5's resource and dependency constraints, and which are fundamentally incompatible with them.

Larger questions remain about whether L5's approach resonates beyond its own design. Will artists, designers, and students find the permacomputing framing compelling or will the friction of a constrained, explicit toolchain outweigh its benefits? Could working within an environment that valorizes older hardware actively influence users to extend the life of their own machines, reducing consumption not just in the tool itself but in the practices it encourages? These questions cannot be answered from within the project alone, and community use, feedback, and research with learners in real contexts will be essential to evaluating whether L5's commitments prove durable in practice.

## 8 CONCLUSION

The five case studies presented here do not resolve the tensions between permacomputing values and the practical demands of a usable creative coding library; they document how those tensions were navigated. Taken together, they suggest that designing sustainable creative tools is less a technical problem than a series of ongoing value judgments: about what to include and what to leave out, what to make visible and what to abstract away, and whose access to prioritize.

Creative coding libraries that consume more system resources and leverage latest technology offer powerful toolkits at the cost of requiring higher specification hardware components and newer computers to keep up with lifecycle changes and graphics requirements. This creates a structural barrier: as libraries grow in capability they narrow the range of machines that can run them. Processing’s Java foundation requires a memory baseline even for any sketch, and p5.js depends on browser rendering engines that grow increasingly resource-intensive with each generation. Game engines like Unity and Unreal Engine, sometimes adopted for creative coding and interactive art, escalate these demands, requiring dedicated graphics hardware, gigabytes of storage for the engine, and frequent updates that make older installations obsolete. Each generation of tooling implicitly assumes a corresponding generation of hardware, compressing the useful lifespan of machines and contributing to cycles of upgrade and disposal. The promise of these more powerful environments is to enable richer real-time graphics, larger datasets, more complex simulations, and a broader feature set that can expand what artists are able to make. Lighter tools preserve access and longevity, but heavier tools can unlock creative possibilities that leaner environments may not be able to easily replicate.

The Processing/p5 paradigm, even when transplanted into Lua, appears to transfer well. Early feedback from usability studies suggest that both beginning creative coders and experienced Processing/p5 users program with L5 with relatively little friction, and that reference examples are accessible and useful. Installation, however, remains a stumbling block: the number of steps required to download, install, and run a first sketch is still higher than ideal, and while video tutorials and step-by-step instructions have helped, this remains a work in progress. Notably, it is worth exploring whether awareness of L5’s low-bandwidth design intent increases adoption interest in L5, as several participants in early usability studies were more willing to work through the additional installation steps and accepting of the underlying stack as they understood the principles underlying design choices in L5. Transparency about values, in other words, proved practically useful, not just philosophically important. Compromises were also necessary in unexpected places. L5 currently uses GitHub as its code forge, reasonable given its ubiquity and the precedent set by other Processing/p5 libraries, but in tension with permacomputing values given GitHub's reliance on extensive cloud infrastructure, its corporate ownership, and its use of code to train LLMs.

No single decision is without cost. Exposing the seams offers more technical understanding, but demands more from users and contributors when they are first learning. Constraining the API preserves longevity but sacrifices familiarity. Designing for older hardware expands access but limits capability. In a resource-constrained future, where continuous production of more powerful machines will be curtailed, designing for older hardware may shift from a principled choice to a practical necessity.

For others considering permacomputing-inspired tools, these trade-offs cannot be avoided, only made deliberately. That means deciding early which principles matter most, who the intended audience is, and where compromises are acceptable. What L5 attempts to demonstrate is that these decisions can be made transparently rather than hidden behind seamless interfaces or deferred to convention. Whether that approach proves durable, for the library, for its users, and for the machines it runs on, remains an open question, and perhaps the most honest measure of whether permacomputing values can take root in creative coding practice.

## ACKNOWLEDGMENTS

The authors wish to thank our reviewers for suggestions that strengthened our arguments. We also wish to thank contributors to the Permacomputing Wiki, the NYC and Berlin Permacomputing Meetups, and contributors to L5 for ongoing discussions that have helped clarify the design decisions documented here.